\documentstyle[aps,epsf,preprint]{revtex}
\begin{document}
\draft
\preprint{Phys. Rev. C (1997) in press.} 
\title{Proton and Antiproton Distributions at Mid-Rapidity
in Proton-Nucleus and Sulphur-Nucleus Collisions}
\author {I.G.~Bearden$^{1}$, H.~B\o ggild$^{1}$, J.~Boissevain$^{2}$,
J.~Dodd$^{3}$, S.~Esumi$^{4,a}$,
C.W.~Fabjan$^{5}$, D.E.~Fields$^{2,b}$, A.~Franz$^{5,c}$,
A.G.~Hansen$^{1}$, E.B.~Holzer$^{5}$, T.J.~Humanic$^{6}$,
B.V.~Jacak$^{2,d}$, R.~Jayanti$^{6,7}$,
H.~Kalechofsky$^{7}$, Y.Y.~Lee$^{7}$,
M.~Leltchouk$^{3}$, B.~L{\"o}rstad$^{8}$,
N.~Maeda$^{4,e}$, A.~Medvedev$^{3}$, 
A.~Miyabayashi$^{8}$, M.~Murray$^{9}$,
S.~Nishimura$^{4,f}$,  S.U.~Pandey$^{7,g}$, F.~Piuz$^{5}$,
V.~Polychronakos$^{10}$, M.~Potekhin$^{3}$,
G.~Poulard$^{5}$, A.~Sakaguchi$^{4,h}$,
M.~Sarabura$^{2}$, M.~Spegel$^{5}$, J.~Simon-Gillo$^{2}$,
W.~Sondheim$^{2}$,
T.~Sugitate$^{4}$, J.P.~Sullivan$^{2}$, Y.~Sumi$^{4}$,
H.~van~Hecke$^{2}$, W.J.~Willis$^{3}$, K.~Wolf$^{9,i}$,
 and N.~Xu$^{2,j}$\\
\center{ (The NA44 Collaboration) \\}}

\address{
$^{1}$ Niels Bohr Institute, DK-2100 Copenhagen, Denmark.\\
$^{2}$ Los Alamos National Laboratory, Los Alamos, NM 87545, USA.\\
$^{3}$ Columbia University, New York, NY 10027, USA.\\
$^{4}$ Hiroshima University, Higashi-Hiroshima 724, Japan.\\
$^{5}$ CERN, CH-1211 Geneva 23, Switzerland.\\
$^{6}$ Ohio State University, Columbus,OH 43210, USA.\\
$^{7}$ University of Pittsburgh, Pittsburgh, PA 15260, USA.\\
$^{8}$ University of Lund, S-22362 Lund, Sweden.\\
$^{9}$ Texas A\&M University, College Station, TX 77843, USA.\\
$^{10}$ Brookhaven National Laboratory, Upton, NY 11973, USA.\\
$^a$ Now at Heidelberg University, Heidelberg, D-69120, Germany\\
$^b$ Now at University of New Mexico, Albuquerque, NM 87185\\
$^c$ Now at Brookhaven National Laboratory, Upton, NY 11973\\
$^d$ Now at State University of New York Stony Brook, Stony Brook, NY 11794\\
$^e$ Now at Florida State University, Tallahassee, FL 32306\\
$^f$ Now at University of Tsukuba, Ibaraki 305, Japan\\
$^g$ Now at Wayne State University, Detroit, MI 48201\\
$^h$ Now at Osaka University, Osaka 560, Japan\\
$^i$ deceased, \\
$^j$ Now at Lawrence Berkeley National Laboratory, Berkeley, CA 94720}
\maketitle

\begin{abstract}

Experiment NA44 has  measured 
proton and antiproton distributions
at mid-rapidity in
sulphur and proton collisions with nuclear targets at 200
and 450 GeV/c per nucleon respectively.
The inverse slopes of transverse mass distributions 
increase with system size for both protons and antiprotons but are 
slightly lower for antiprotons.
 This could happen if antiprotons are annihilated
in the nuclear medium. 
The antiproton yield increases with system size and centrality
and is largest at mid-rapidity. The proton yield 
also increases with system size and centrality, but decreases from backward 
rapidity to mid-rapidity. 
The stopping of protons at these energies
lies between the
full stopping and nuclear transparency scenarios.
The data are in reasonable agreement
with RQMD predictions except for the antiproton yields from sulphur-nucleus
collisions.
\end{abstract}
 

\section{Introduction}

  Nucleus-nucleus collisions at ultrarelativistic energies
create hadronic matter at high energy density. 
The distributions of baryons at mid-rapidity provide a sensitive probe 
of the collision dynamics.
 In particular, the
stopping power determines how much of the incoming longitudinal
energy is available for excitation of the system.
These collisions
have been described by microscopic models incorporating hadron
production and rescattering, see for example \cite{Sor89a}. 
The data reported here impose constraints on the amount of stopping
of baryons in such models.

Enhanced production of antibaryons may 
indicate formation of a state of matter in which 
the quarks and gluons are deconfined
\cite{Hei84a,Koc88a,Ell89a,Lee88a}. Such
enhancement may be hidden, however, by antibaryon annihilation 
with baryons  
 \cite{Gav90a,Sor90a}; the antiproton survival 
probability is sensitive to both the collision environment and the
antiproton formation time. 
The 
antiproton and proton distributions may
also reflect the degree of thermalization achieved and,
by comparing distributions
from light and heavy systems,
allow detailed studies of rescattering.

We present proton and antiproton measurements using the NA44
spectrometer
from $pBe$ (to approximate $pp$), proton-nucleus and 
nucleus-nucleus interactions. This allows a systematic study 
as a function of the size of the central
region and different conditions in the surrounding hadronic matter.
 
These systematic studies are aided by use of an event generator. 
The RQMD model, version 1.08 \cite{Sor89a},  is a 
microscopic phase space approach, based on resonance and string
excitation and fragmentation with subsequent hadronic collisions.
RQMD includes annihilation of antiprotons in the hadronic
medium when they collide with baryons \cite{Sor90a}.
We study a feature of RQMD called color ropes by the authors of the
model. RQMD uses a string model of particle production from
each nucleon-nucleon collision. In a heavy-ion collision, where
there are numerous nucleon-nucleon collisions, the density of these 
strings is high and multiple strings overlap. Overlapping strings
do not fragment independently but form `ropes',
chromoelectric flux-tubes whose sources are 
color octet charge states
rather than the color singlet charges of normal strings \cite{Sor92a}.
These ropes represent a collective effect in nucleus-nucleus collisions,
and have been shown to enhance both strangeness and 
baryon pair production \cite{Sor92a}.

\section{Experiment}

\begin{figure}
     \epsfxsize=16cm
     \epsffile{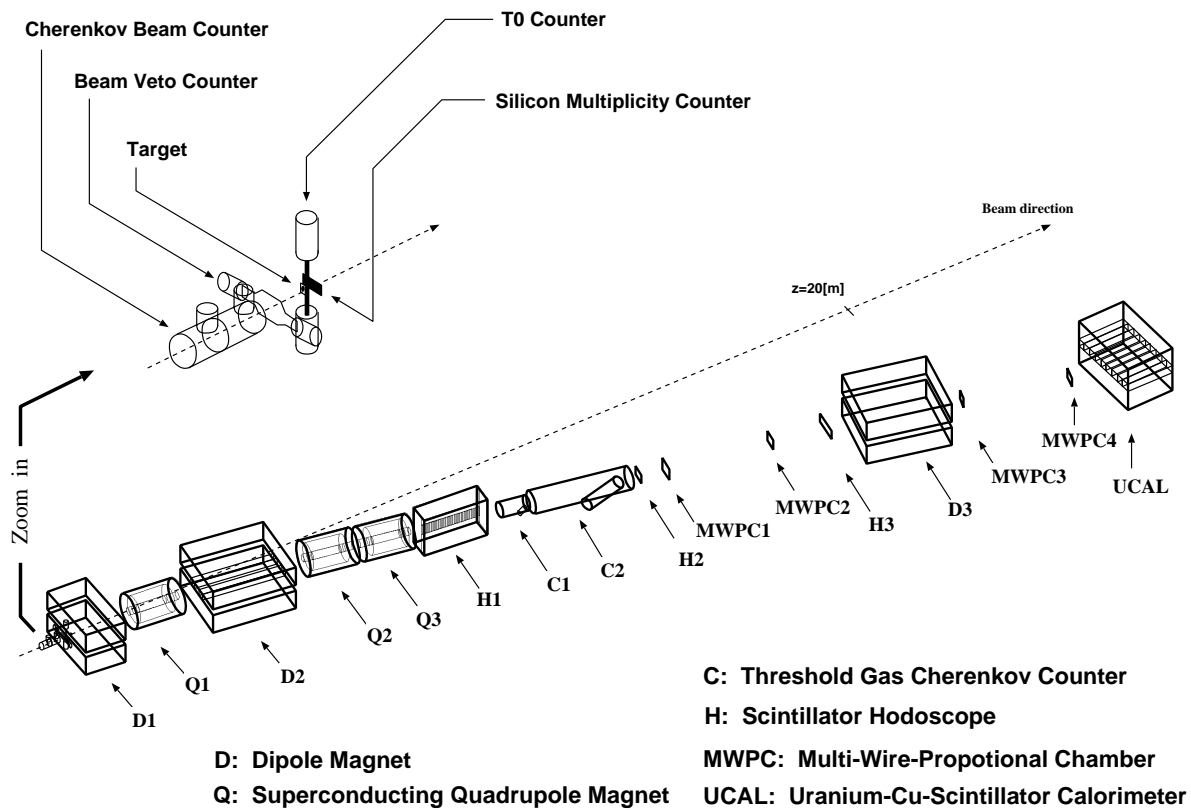}
    \caption{The NA44 experimental set-up.}
    \label{fg:setup}
\end{figure}

\begin{figure}
    \begin{center}\mbox{  
        \epsfxsize=16cm
        \epsffile{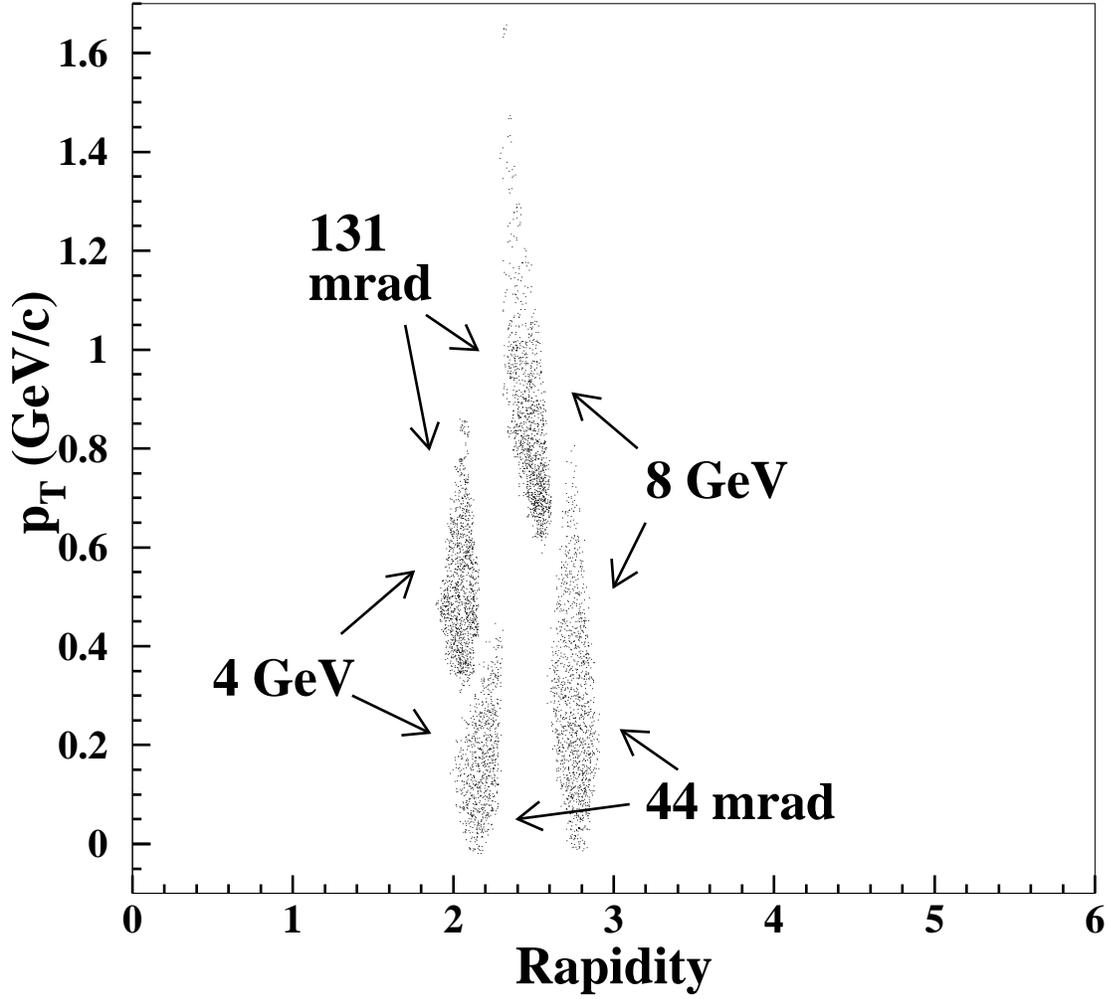}}\end{center}
    \caption{The $p$ or $\bar{p}$ acceptance in y and $p_{T}$.
The $\phi$ acceptance (not shown) decreases from
$2\pi$ at $p_{T}$~=~0 to 0.012 at $p_{T}$~=~1.6~GeV/c.}
    \label{fg:acprot}
  \end{figure}

The NA44 experiment is shown in Figure \ref{fg:setup}.
Three conventional dipole magnets (D1, D2, and D3) and three 
superconducting quadrupoles 
(Q1, Q2, and Q3) analyze the momentum and create a magnified 
image of the target in the spectrometer. The magnets focus particles 
from the target onto the first hodoscope (H1) such that
the horizontal position along the hodoscope gives the total momentum.
Two other hodoscopes (H2, H3) measure the angle of the 
track.
The momentum acceptance 
is $\pm$ 20\% of the nominal momentum setting. 
The angular coverage is  approximately -5 to +78 mrad
with respect to the beam in the
horizontal plane and $\pm$ 5 mrad vertically.
Only particles of a fixed charge sign are detected in a given 
spectrometer setting.
Four settings are used to cover the mid-rapidity 
region in the $p_T$ range 0 to 1.6 GeV. 
Figure \ref{fg:acprot}
shows the acceptance of the spectrometer in the $y-p_T$
plane for the 4 and 8 GeV/c momentum settings when the 
spectrometer axis is at  44 and 131 mrad with respect to the beam.
 
For the sulphur-nucleus data, the beam rate and time-of-flight start are
determined using a Cherenkov beam counter (CX), with time resolution
of approximately 35~ps \cite{NA4492c}. 
For the proton-nucleus data,
a scintillator counter is used to measure the beam rate.
A second scintillator (T$_0$) is used to trigger on central events in
sulphur-nucleus collisions by requiring
a large pulse height (high charged particle multiplicity). 
The pseudorapidity coverage of T$_0$ is roughly 1.3 to 3.5.
For proton beams, T$_0$ provides
the interaction trigger by requiring that at least one charged particle hit
the scintillator, and also provides the time-of-flight start with a 
time resolution of approximately 100~ps. 
A silicon pad detector measures the 
charged-particle multiplicity with $2 \pi$ azimuthal acceptance in the 
pseudorapidity range $1.8<\eta<3.3$.

\begin{figure}
    \begin{center}\mbox{  
        \epsfxsize=16cm
        \epsffile{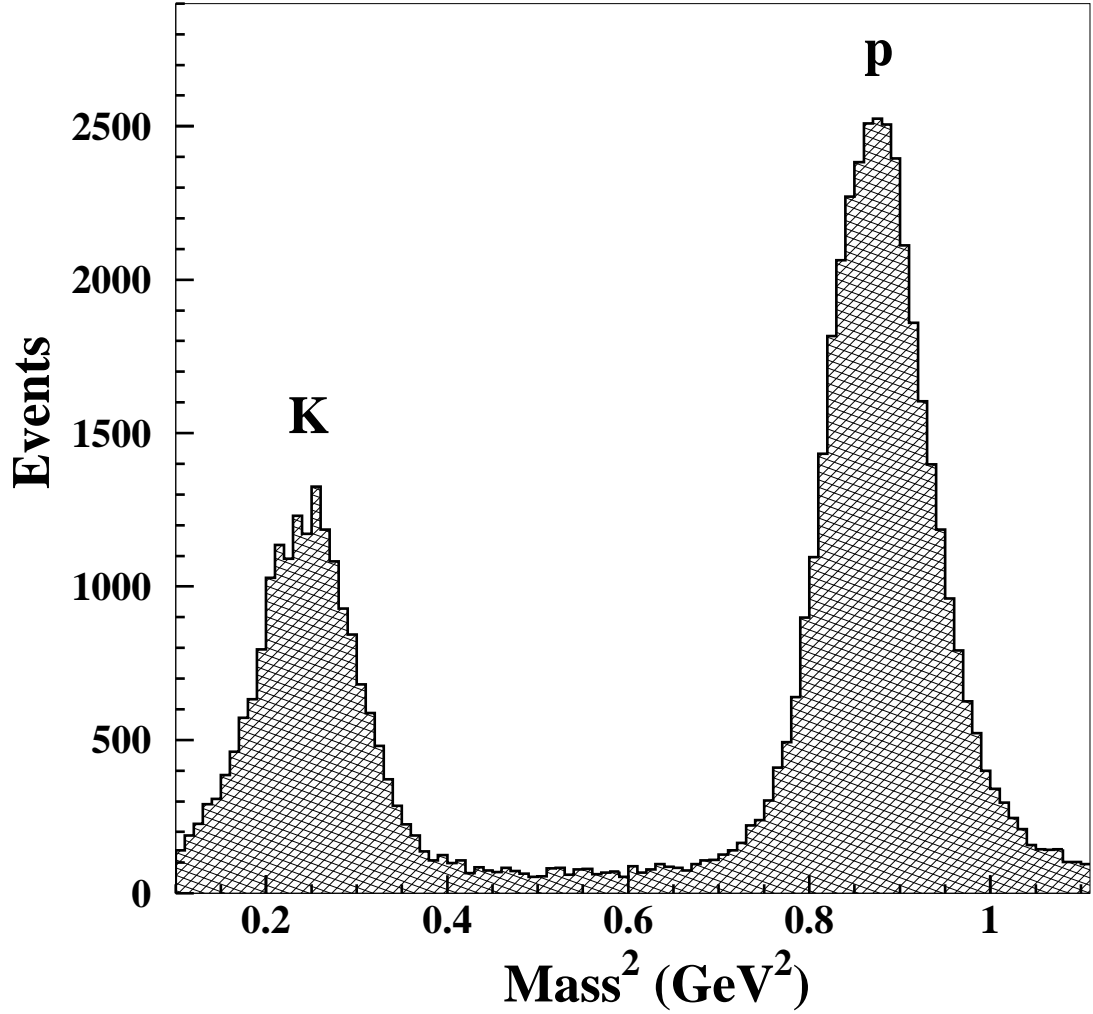}}\end{center}
    \caption{Mass-squared distribution from Hodoscope 3 after pions have been
   vetoed by the Cherenkovs.}
    \label{fg:pidprot}
  \end{figure}

The three scintillator hodoscopes (H1, H2 and H3) are used to track
the particles and are divided into 50, 60 and 50 slats, respectively. 
The hodoscopes also provide time-of-flight with a time resolution
of approximately 100~ps; particle identification relies primarily
upon the third hodoscope.
Two Cherenkov counters differentiate kaons and protons
(C1: freon-12 at 1.4 or 2.7 atm, depending on the spectrometer setting), 
and reject electrons and pions
(C2: nitrogen/neon mixture at 1.0 or 1.3 atm).  
An appropriate combination of C1 and C2 is used for each
spectrometer setting to trigger on events with no pions or
electrons in the spectrometer.
Particles are identified by their 
time-of-flight,
in combination with the Cherenkov information.
Figure \ref{fg:pidprot} illustrates the particle identification after pions 
have been vetoed by the Cherenkovs; kaons and protons are clearly separated.
More details about the spectrometer are available in \cite{NA44mt}.
 
\section{Data Analysis}

 The proton and antiproton data samples after particle identification 
and quality cuts are shown in Table~\ref{tb:data}.
Also shown for each data set is the target 
thickness and the centrality, expressed as a fraction of 
the  total inelastic cross-section. 

Tracks are reconstructed from the hit positions on the three
hodoscopes, 
constrained by straight-line 
trajectories after the magnets.
In order to construct the invariant cross-section,
the raw distributions are corrected using 
a Monte Carlo 
simulation of the detector response. Simulated tracks are passed 
through the full analysis software chain and used to correct the data 
for geometrical acceptance, reconstruction efficiency and momentum 
resolution. Particles are generated according to an exponential
distribution in transverse mass, $m_T=\sqrt{p_T^2+m_p^2}$,
with the coefficient of the exponent determined
iteratively from the data.

 The Cherenkovs reject pions with an efficiency
of greater than 98\% at the trigger level.
Further offline selection reduces the pions
to a few percent of the kaons.
After time-of-flight and Cherenkov cuts,
the residual kaon contamination of the proton sample 
is less than 3\%. 

The invariant cross-sections are presented as a function of 
$m_T$ - $m_p$, where $m_p$ is the mass of the proton. 
The absolute normalization of each spectrum is calculated
using the number of beam particles,
the target thickness, the fraction of interactions satisfying 
the trigger, and the measured live-time of the data-acquisition
system. For the $SA$ data, the centrality
selection is determined by comparing the pulse height
distribution in the T$_0$ counter 
for central and unbiased beam triggers.
For $pA$ systems, the 
fraction of inelastic collisions
producing at least one hit in the interaction (T$_0$)
counter is modeled with the event
 generators RQMD \cite{Sor89a} and FRITIOF \cite{FRIT87,FRIT92}.
The errors on the centralities for the $pA$ data in Table~\ref{tb:data}
reflect the systematic uncertainty on this fraction from comparing the
two models.   
The $pA$ data are also corrected for the efficiency of the
T$_0$ counter.
The resulting centrality fractions are indicated in 
Table~\ref{tb:data}. 

The proton-nucleus data are  corrected for non-target background.
The largest corrections are 7.4\% and 6.7\% 
 on the absolute cross-sections for protons and antiprotons
from $pBe$ collisions. This correction does not affect the shape 
of the distribution. 
The corrections to the nucleus-nucleus data are negligible.
The cross-sections are also corrected for
the proton identification efficiency,
and for the effects of
selecting events with no accompanying pion or electron.
The pion (electron) veto correction is determined from the number of protons
in runs for which the pions (electrons) are not vetoed.
The effect of 
hadronic interactions of the produced particles in the material of the
spectrometer has been studied using a detailed GEANT simulation. 
These interactions, including annihilation of  antiprotons in the
spectrometer material, do  
not distort the shape of the measured transverse mass distributions
but result in a reduction in the observed yields
of about 11\% for protons and  17\% for antiprotons.
The data are corrected for these losses.

The invariant cross-sections, measured in the NA44 acceptance, are
generally well described by exponentials in transverse mass (see Equation
\ref{eq:mt}).
The proton and antiproton rapidity densities (dN/dy)
are calculated by integration of the normalized $m_T$ 
distributions, 
with the fitted coefficient of the exponent (the `inverse slope') 
used to extrapolate to high $m_T$, beyond the region of measurement. 
The statistical error on this extrapolation
is calculated using the full error matrix from the fit of 
Equation \ref{eq:mt} (see Section~\ref{sec:results}) to the $m_T$ spectrum.
The corresponding systematic error is included in Table~\ref{tb:errors}.

\section{Systematic Errors}

The momentum scale of the spectrometer is verified with a second, independent
measurement of the momentum using
the multiwire proportional chambers (MWPCs 1-4 in Figure \ref{fg:setup}) 
and dipole magnet (D3),
yielding a systematic 
error of 1.6\% on the $p_T$ scale. 
Since the NA44 spectrometer has acceptance for both
positive and negative $p_X$, 
a systematic offset in
$p_T$  can be checked by requiring symmetry around
$p_X=0$. The resulting uncertainty on the origin of
 the $p_T$ scale is 7 MeV/c for the 8~GeV setting. 
A detailed Monte Carlo simulation, including multiple scattering and detector
granularity, is used to correct for the finite resolution of the spectrometer,
and introduces a systematic error in $p_T$ of 0.15\%.

Systematic errors on the inverse slopes of the transverse mass distributions
are estimated by comparing the inverse slope determined from the 131 mrad 
data to the inverse slope determined from both angle settings, 
and are less than 5\%(15\%) for the 8~GeV(4~GeV) setting data.
The systematic errors 
due to the spectrometer acceptance correction are
estimated from the sensitivity of the extracted slope to the fit
range used,
and by measuring the slope determined from
the ratio of the cross-sections 
corresponding
to the `central ray' of the spectrometer 
at both the 131 and 44 mrad settings.
In this ratio the acceptance corrections cancel since the particles
have the same path through the spectrometer. 
The total uncertainty
is 10 MeV/c for the mid-rapidity ($y$=2.3-2.9) 
data, and 10-20 MeV/c for the lower rapidity ($y$=1.9-2.3) data. 

The error in the absolute normalizations is dominated
by the uncertainty in the fraction of the total cross-section 
selected by the NA44 centrality trigger, and by the pion and 
electron veto corrections. The relative error in the
centrality is 6\% for both the $SS$ and $SPb$ data, resulting in a
systematic uncertainty of 6\% in dN/dy. 
For the
proton-nucleus data, the trigger bias is determined by modeling
the acceptance of the T$_0$ counter, as described above.
The resulting dN/dy values
are sensitive to the charged particle distribution from the two
models, giving an uncertainty of 1.5\% for the $pBe$ data,
increasing to 3\% for $pPb$. 
 Corrections for the fraction of events vetoed by pions are significant
only for 
the sulphur-nucleus data at the 44 mrad setting.
 The uncertainties in these corrections  are 10\% for 
$SPb$ collisions and 5\% for $SS$ collisions.
Additional errors on the dN/dy values arise from a 5\%
uncertainty in the determination of the data-acquisition
live-time, and a 5\% beam rate
dependence of the pion veto correction for the $SPb$ data.
For the $pA$ data, there is an additional 6\% error due to
the correction for the efficiency of the interaction counter.
The correction for non-target background contributes a negligible
 systematic error to the absolute cross-sections for proton-nucleus data. 
The total systematic errors on the measured inverse slopes and
dN/dy values are given in Table~\ref{tb:errors}.

\section{Feed-Down from Weak Decays}

Weak decays of strange baryons are a significant source 
of protons and antiprotons, and contribute to the yields
measured in the NA44 spectrometer.
A strange baryon may travel several centimeters from the target before
decaying weakly to a proton and a pion. Such a proton
may be reconstructed in the spectrometer with the wrong momentum.
The sensitivity of the data to this feed-down has been studied
using a GEANT simulation of the spectrometer with particle
distributions and yields taken from RQMD.
Of the strange baryons, only $\Lambda^0$ and $\Sigma^+$ 
decay weakly to protons (with the corresponding antiparticles  
decaying to antiprotons).
Heavier strange baryons also contribute via sequential decays.

\begin{figure}
    \begin{center}\mbox{  
        \epsfxsize=16cm
        \epsffile{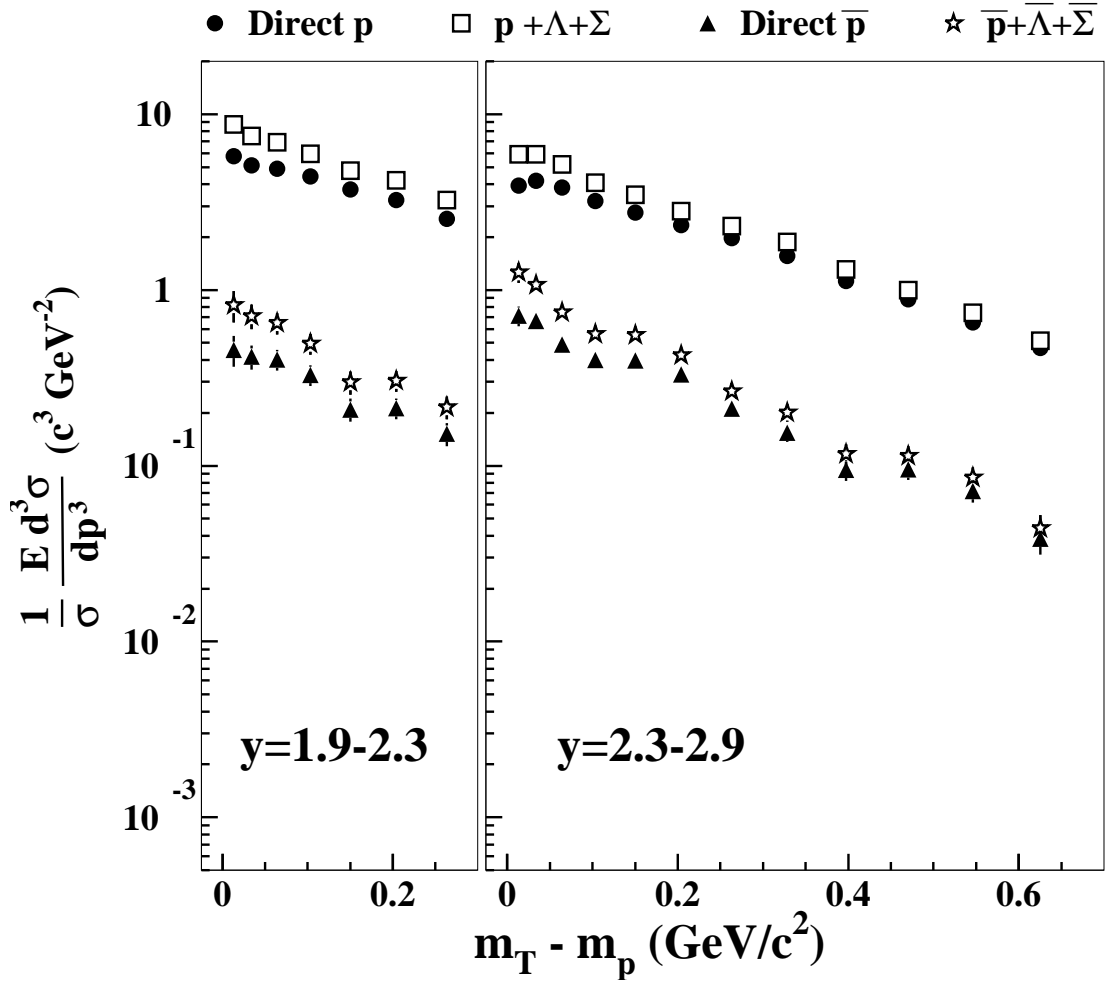}}\end{center}
    \caption{Invariant cross-sections as a function of $m_T - m_p$ 
for central $SPb$ collisions from RQMD,
with (open symbols) and without (solid symbols) feed-down from weak decays.}
    \label{fg:rqmdfeed}
\end{figure}

The fraction of the measured protons (antiprotons) arising from these
decays is calculated as a function of rapidity and transverse momentum
within the NA44 acceptance, 
and used to determine 
`feed-down factors' for the $m_T$ and dN/dy distributions.
These are multiplicative factors which provide an estimate of the contribution
of $\Lambda$ and $\Sigma$ decays to the measured distributions, 
and could be applied to the numbers in Tables~\ref{tb:slope} and \ref{tb:dndy}
to estimate the inverse slopes and yields of `direct' protons (antiprotons).
Figure \ref{fg:rqmdfeed} shows the effect of feed-down from weak decays 
on the RQMD proton and antiproton transverse mass distributions for 
$SPb$ collisions.
Including $\Lambda$ and $\Sigma$ decays tends to make the distributions
steeper since the protons arising from weak decays contribute more
at low $p_T$.
Table~\ref{LAMCOR} shows the
feed-down factors for the inverse slopes and 
yields of the data
due to $\Lambda$ and $\Sigma$ decays.
The values are calculated 
using the $\Lambda/p$ and $\Sigma/p$ ratios from RQMD.
The `errors' reflect the result of 
increasing and decreasing the $(\Lambda +\Sigma)/p$ ratio in the model by
a factor of 1.5.
This factor of 1.5 is consistent with the scale of the 
discrepancies in the 
published data on $\Lambda$ production in nucleus-nucleus collisions.
For central $SS$ collisions, RQMD is consistent with the $\Lambda/p$
ratio measured by NA35 \cite{S95NA35}. However for $SPb$,
there is a factor of 2 discrepancy
between the measurement of  $\Lambda$ production by NA36 \cite{S95NA36} 
and the scaled $SAg$ data of NA35.
The RQMD prediction lies between the results of the two experiments.
The feed-down factors have been calculated explicitly for protons from 
$pBe, pPb$ and $SPb$
using the complete GEANT simulation, and are scaled 
according to the respective RQMD ($\Lambda +\Sigma)/p$ ratios
for protons from $pS$ and $SS$, and for antiprotons from all systems.
A factor of 1.5 variation in the ($\bar{\Lambda} + \bar{\Sigma})/\bar{p}$
ratio is also assumed in calculating the antiproton factors.
The $m_T$ dependence of the feed-down factors is 
mainly determined by the experimental acceptance and not by the $m_T$ 
dependence of the ($\Lambda +\Sigma)/p$ ratio from RQMD.

As these factors necessarily contain some 
dependence on models they have
not been applied to the data but are listed 
in Table~\ref{LAMCOR}
so that the reader 
can appreciate the sensitivity of the data to 
$\Lambda$ and $\Sigma$ decays.

\section{Results}
\label{sec:results}

\begin{figure}
    \begin{center}\mbox{  
        \epsfxsize=16cm
        \epsffile{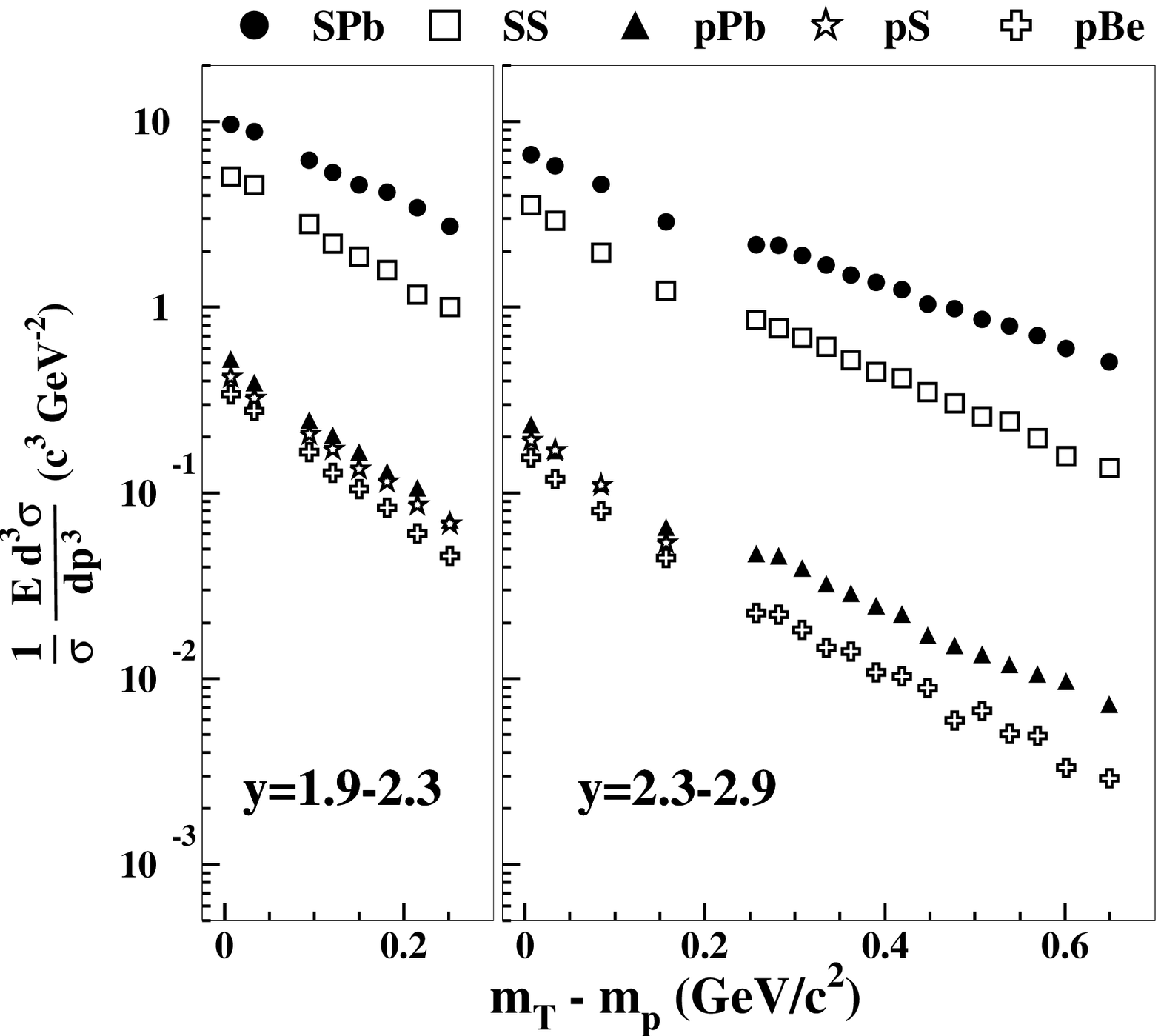}}\end{center}
    \caption{Invariant cross-sections for protons 
as a function of $m_T$ - $m_p$.}
    \label{fg:prot48}
\end{figure}

\begin{figure}
    \begin{center}\mbox{  
        \epsfxsize=16cm
        \epsffile{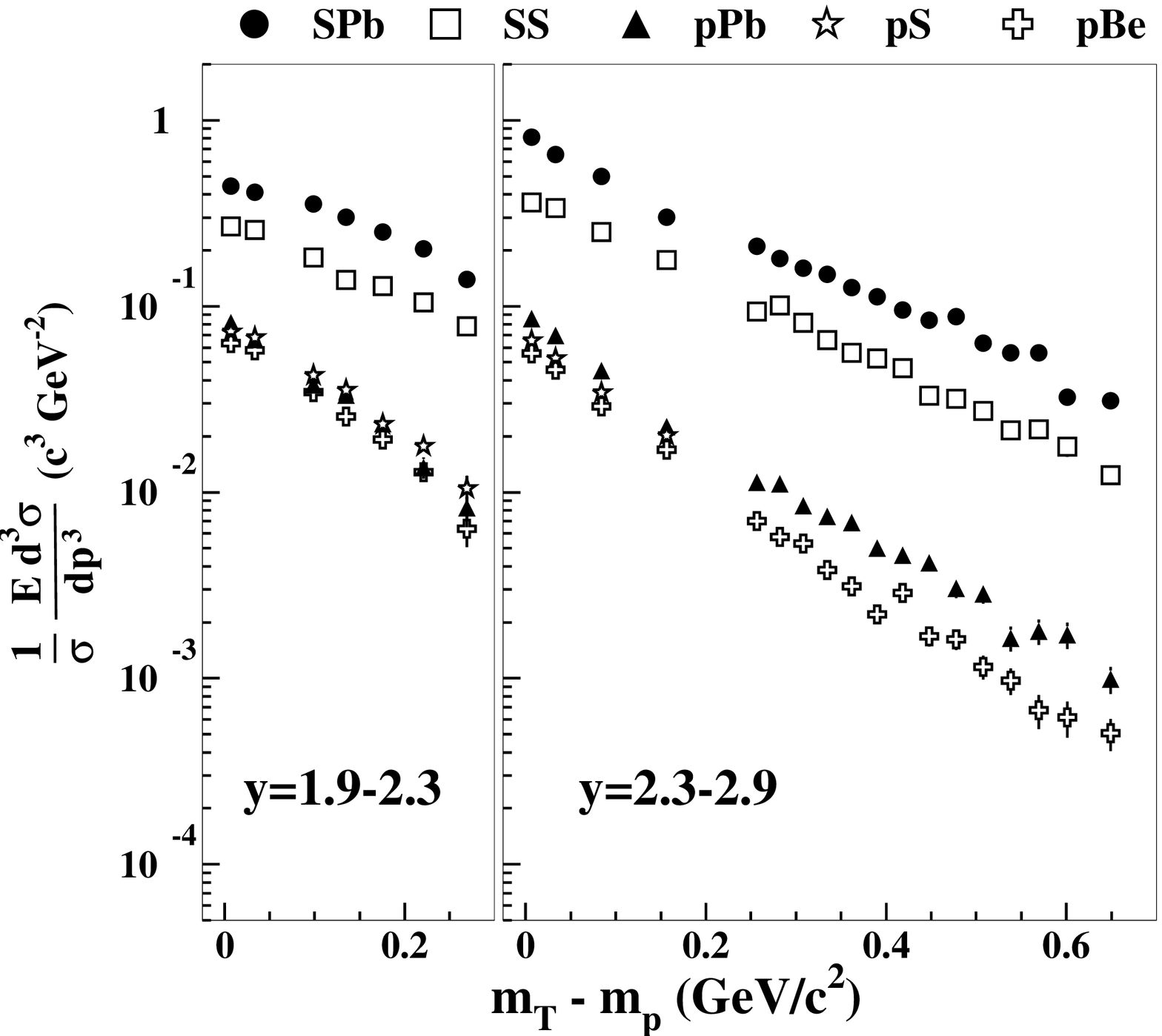}}\end{center}
    \caption{Invariant cross-sections for antiprotons 
as a function of $m_T$ - $m_p$.}
    \label{fg:pbar48}
  \end{figure}

\begin{figure}
  \begin{center}\mbox{  
  \epsfxsize=16cm
  \epsffile{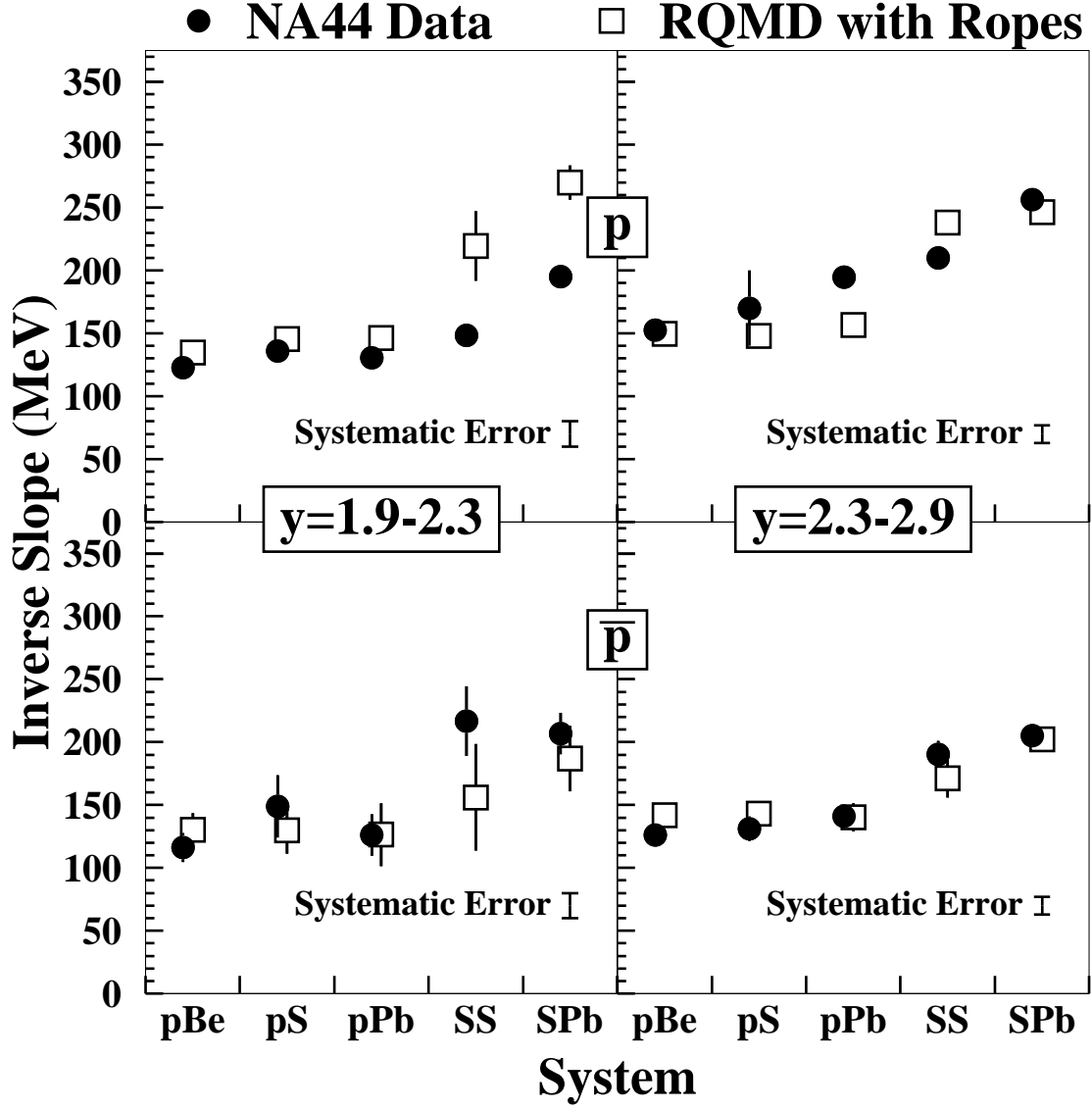}}\end{center}
  \caption{Inverse slopes of the transverse mass distributions for
    each system for
    data and RQMD. Statistical and systematic errors for the data
    are added in quadrature. The global systematic errors common
    to all systems (Table~\protect{\ref{tb:errors}})
    are shown by bars near 
the bottom right hand corner of each plot}
  \label{fg:tvrqmd}
\end{figure}

The invariant cross-sections for protons and antiprotons from
$pBe$, $pS$, $pPb$, $SS$  and $SPb$ collisions are shown
in Figures~\ref{fg:prot48} and \ref{fg:pbar48} as a function of 
$m_T$ - $m_p$. 
The transverse
mass distributions are generally well described by exponentials in the 
region of the NA44 acceptance:
\begin{equation}
 \frac{1}{\sigma}\frac{E d^{3}\sigma}{dp^{3}} = Ce^{-(m_T-m_p)/T}
\label{eq:mt}
\end{equation}
where C is a constant and  $T$ the inverse logarithmic slope.
The inverse slope parameters obtained by fitting the 
proton and antiproton data to Equation~\ref{eq:mt} 
are given in Table~\ref{tb:slope},
and plotted in Figure~\ref{fg:tvrqmd}. 
The inverse slopes for both protons and antiprotons increase with system
size. 
For protons, the inverse slopes are higher at mid-rapidity ($y$=2.3-2.9)
than at more backward rapidities ($y$=1.9-2.3).
This effect is not seen for antiprotons, where the inverse slopes are similar
in the two rapidity intervals. Note that the errors on the backward rapidity
data are significant.
The inverse slopes for antiprotons are
generally somewhat lower than for protons at mid-rapidity, but are comparable
in the backward rapidity interval.

\begin{figure}
  \begin{center}\mbox{  
  \epsfxsize=16cm
  \epsffile{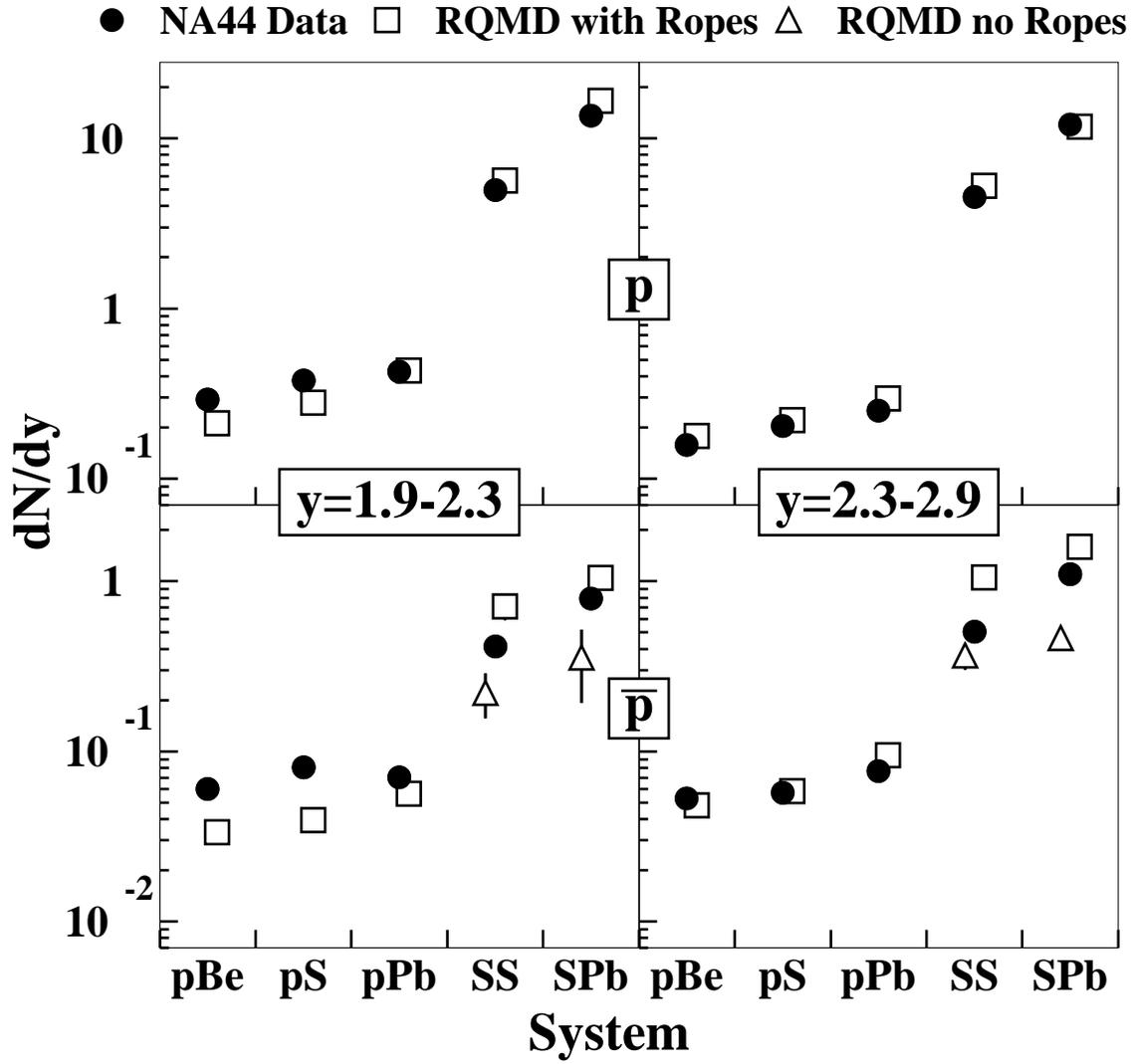}}\end{center}
  \caption{Rapidity densities (dN/dy) for
    each system for
    data and RQMD. Statistical and systematic errors for the data
    are added in quadrature.}
  \label{fg:dnvrqmd}
\end{figure}

The proton and antiproton rapidity densities (dN/dy)
are listed in Table~\ref{tb:dndy} and plotted 
in Figure~\ref{fg:dnvrqmd}. 
Proton yields increase 
with system size, and are significantly larger in 
nucleus-nucleus collisions than in proton-nucleus collisions.
More protons are produced in the backward rapidity interval ($y$=1.9-2.3)
than at mid-rapidity ($y$=2.3-2.9), particularly for  
proton-nucleus collisions.
The antiproton yields are lower than the proton yields and grow less rapidly
with increasing system size:
the increase in the antiproton yield between $pBe$ and $pPb$ is less than 
50\%.
Comparing antiproton yields in the two rapidity intervals, there is essentially
no difference for proton-nucleus collisions. 
In nucleus-nucleus collisions however, antiproton production is
notably smaller backwards of mid-rapidity.

\begin{figure}
    \begin{center}\mbox{  
        \epsfxsize=16cm
        \epsffile{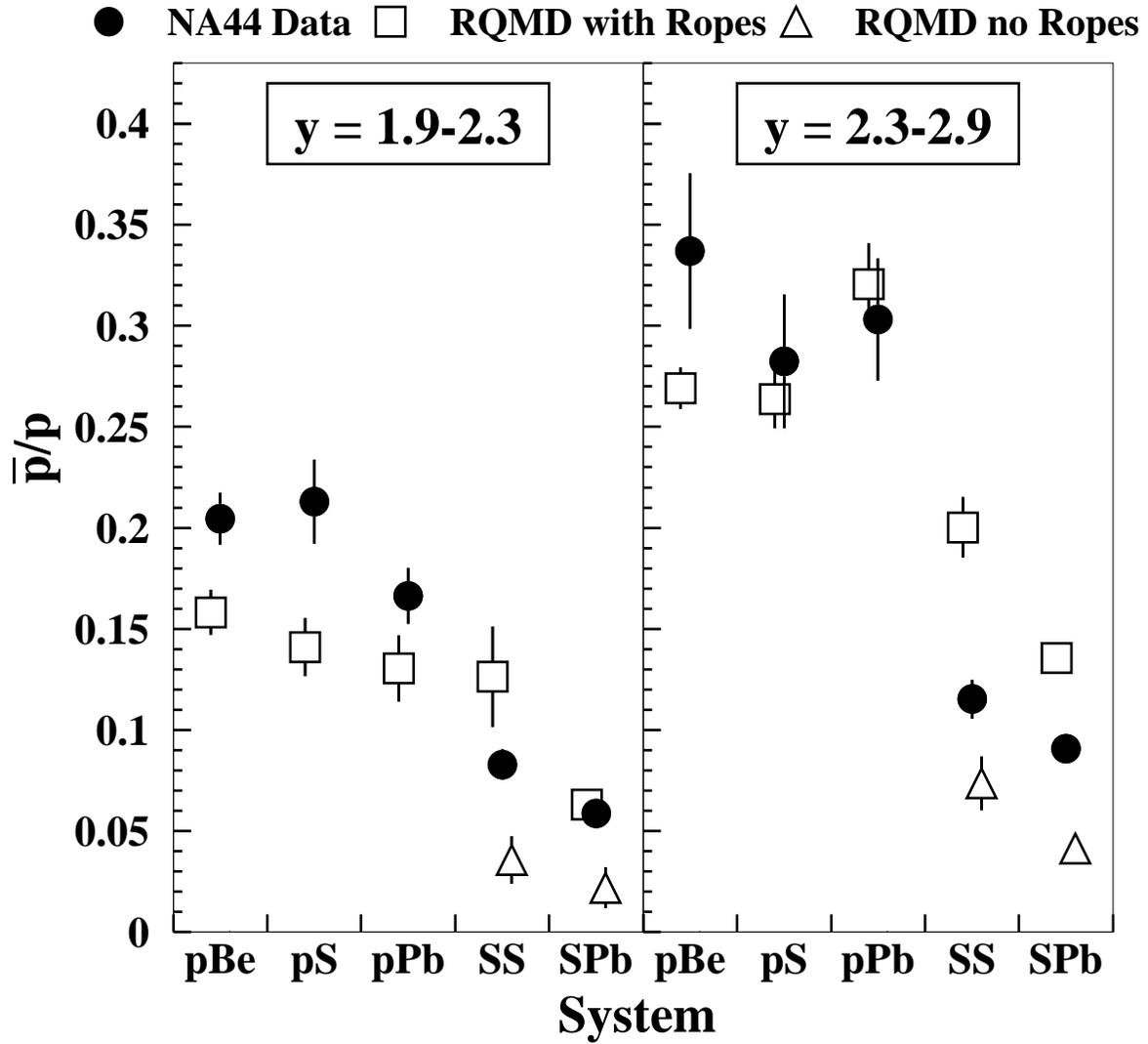}}\end{center}
    \caption{The integrated $\bar{p}$ to $p$ ratio for data and RQMD.
 Systematic and statistical errors have been added in quadrature. 
 The value of the integrated ratio for 
 $pp$ collisions \protect{\cite{Gue76a}} is $0.33 \pm 0.13$.}
    \label{fg:ratio}
 \end{figure}

Figure~\ref{fg:ratio} shows the ratio of $\bar{p}$ to $p$ yields 
for the various projectile-target systems.
Note that the systematic errors described in Table~\ref{tb:errors} cancel 
in this ratio. 
The $\bar{p}/p$ ratio decreases by 
a factor of 4 from $pBe$ to $SPb$. Most of this decrease with
system size occurs between $pPb$ and $SS$. The target dependence of the
ratio is stronger in sulphur-nucleus than in proton-nucleus collisions.
Comparing the two rapidity intervals, the $\bar{p}/p$ ratio is larger at
mid-rapidity in all cases. 

The beam momentum for the $pA$ data is 450 GeV/c, 
corresponding to a beam rapidity of 7, 
while for the $SA$ data the beam momentum is 200 GeV/c per nucleon,
corresponding to a beam rapidity of 6.
The energy dependence of proton and antiproton production 
in $pp$ collisions has
been studied at the ISR \cite{Gue76a}. 
Using the parametrizations of the proton and antiproton cross-sections 
as a function of center-of-mass energy from \cite{Gue76a}, 
the effect of the different beam momenta for the 
$pA$ and $AA$
data on the systematic behaviour of the $\bar{p}/p$ ratio 
can be estimated.
Decreasing the beam momentum from 450 to 200 GeV/c per nucleon,
this energy scaling implies that the $\bar{p}$ yield decreases by 
$12\pm1\%$ and the $p$ yield decreases by $1.5\pm0.1\%$.
Thus the decrease in beam energy cannot explain the decrease of the 
$\bar{p}/p$ ratio. Rather it reflects the fact that 
at mid-rapidity most protons are 
not produced but originate from the target or
projectile. 

\begin{figure}
 \begin{center}
    \mbox{
     \epsfxsize=16cm
     \epsffile{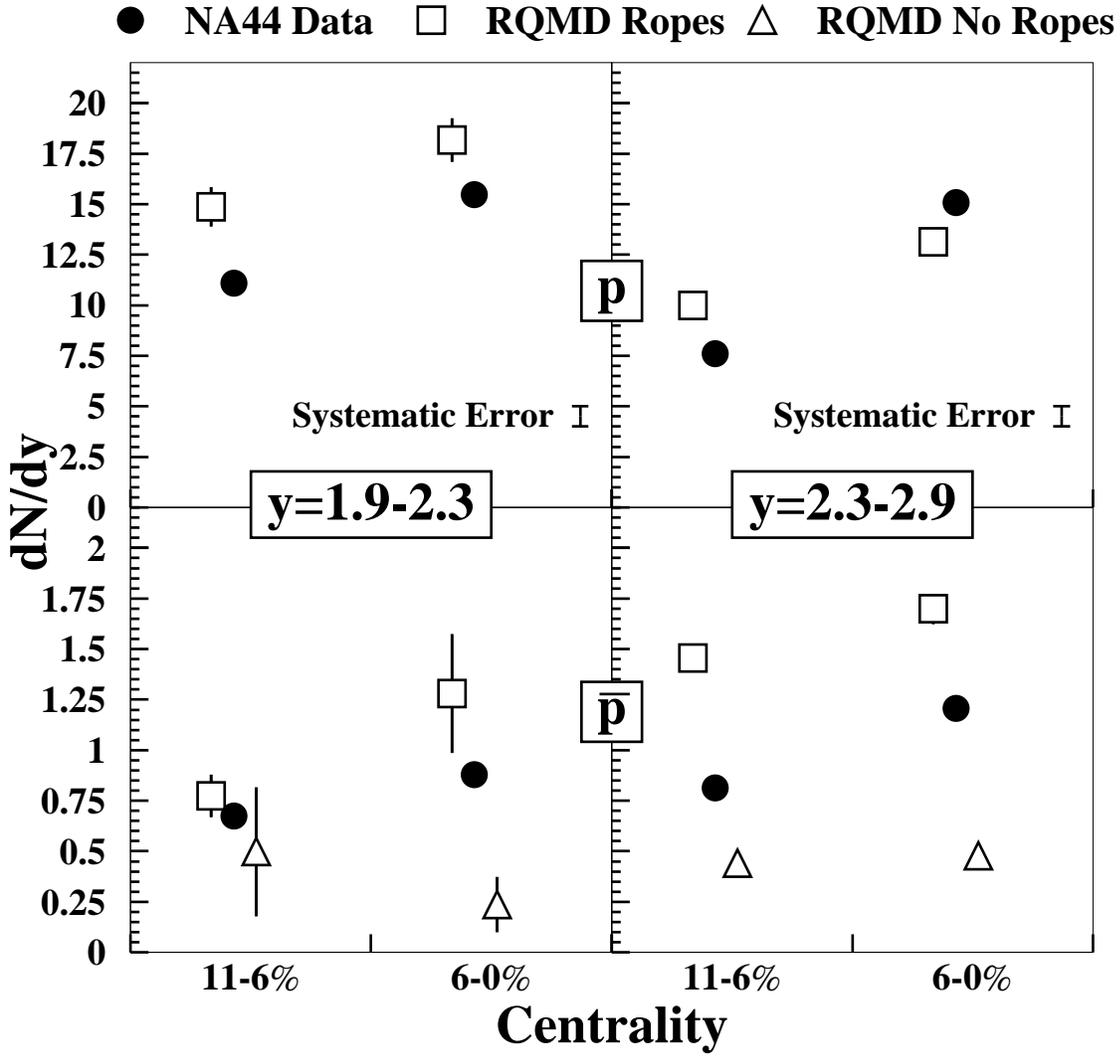}
      }
  \end{center}
  \caption{Proton and antiproton yields as a function of
centrality for $SPb$ collisions.
Statistical errors are shown for each point.
The vertical bar shows the systematic error of 13.6\%
common to all of the data, see Table~\protect{\ref{tb:errors}}.
Also shown are the RQMD predictions for the same centrality
selection.}
 \label{spbvcent}
\end{figure}

In order to study the centrality dependence of proton and antiproton 
production, the $SPb$ data are divided into two 
subsamples containing the 
11-6\% and 6-0\% 
most central collisions
respectively.
The inverse slopes show no centrality dependence between the two
subsamples.
Figure \ref{spbvcent} shows 
the proton and antiproton yields for these two different centrality 
selections.
Production of both protons and antiprotons increases with centrality,
although the yield of protons rises faster.

\section{Discussion}

\begin{figure}
    \begin{center}\mbox{  
        \epsfxsize=16cm
        \epsffile{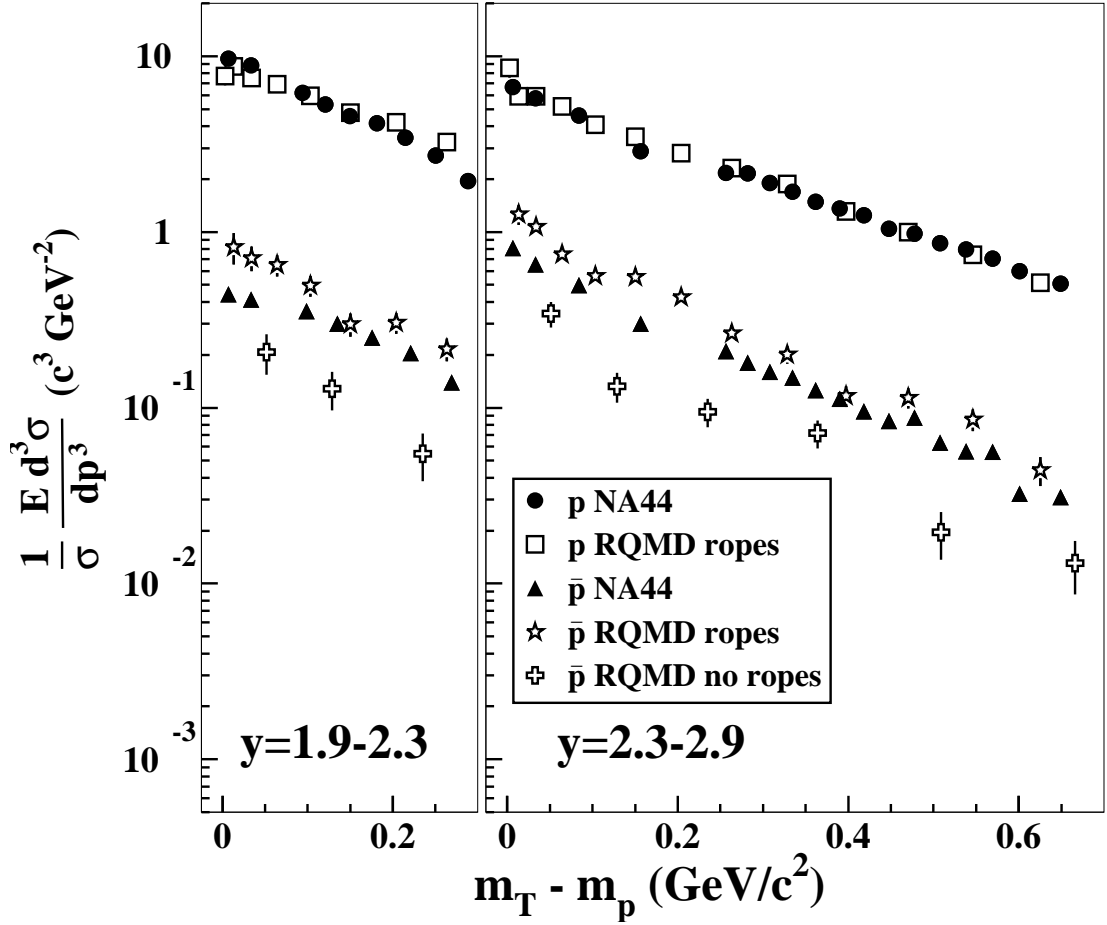}}\end{center}
    \caption{Invariant cross-sections as a function of  $m_T - m_p$ 
for central $SPb$ collisions from data and RQMD.
The RQMD distributions include the effect of feed-down from weak decays.}
    \label{fg:spbvrqmd}
\end{figure}

The precision of the data and the range of the systems studied 
 provide strong constraints on models of proton and 
antiproton production,
rescattering and annihilation.
We have compared our data in detail to the RQMD model.
Figure~\ref{fg:spbvrqmd} compares the transverse mass distributions 
of protons and antiprotons from $SPb$ collisions 
to predictions from RQMD with and without rope formation. 
The contribution of feed-down from $\Lambda$ and $\Sigma$ decays
is included in the model predictions, determined using the GEANT
simulation described above.
The shapes of the spectra are generally reproduced by the model.
This is true also for the lighter collision systems.
The RQMD distributions are then used to determine inverse slopes and yields
from the model in the same way as for the data.

Comparing different collision systems, the inverse slopes of both protons 
and antiprotons increase with system size. 
The relative increase, from $pBe$ to $SPb$, is similar in both cases.
NA44 has previously reported an increase in the inverse slopes of kaons, 
protons and antiprotons produced at mid-rapidity in symmetric collision
systems \cite{NA44pbpb}. These data extend this trend to asymmetric systems,
and to more backward rapidities ($y$=1.9-2.3).
For symmetric systems, the inverse slopes of protons and antiprotons are
equal, while for the asymmetric systems $pS, pPb$ and $SPb$ at $y$=2.3-2.9
the inverse slopes of protons are higher than those of antiprotons.

The increase of the proton inverse slope
with system size, which continues 
through $PbPb$ collisions \cite{NA44pbpb}, is
a result of the increasing number of produced
particles and consequent rescattering.
The large number of secondary collisions causes the hadronic
system to expand \cite{NA44mt,NA44pbpb,fields}, and the velocity boost from
this collective expansion is visible as an increased inverse slope in the 
proton $m_T$ distributions.  One would expect this effect to be 
concentrated at mid-rapidity, and this is supported by the data which
show that the 
inverse slopes of protons are higher at $y$=2.3-2.9 than at $y$=1.9-2.3.
Though the antiproton inverse slope is also
increased by secondary collisions, the RQMD model predicts that
the observed antiprotons have suffered fewer rescatterings than
the observed protons. This is because secondary collisions
with baryons can annihilate the antiprotons, and the number of
baryons at mid-rapidity in the nucleus-nucleus collisions is
substantial \cite{NA44pbpb}. 
This may explain why the inverse slopes of antiprotons 
do not increase from $y$=1.9-2.3 to $y$=2.3-2.9.

RQMD reproduces the trend of increasing inverse slope with system size
(Table~\ref{tb:slope} and Figure~\ref{fg:tvrqmd}), 
and is in reasonable agreement on the
absolute values of the inverse slopes, with the exception of the backward
rapidity ($y$=1.9-2.3) protons from sulphur-nucleus collisions, 
where the model predicts larger inverse slopes than are measured 
experimentally.
For $SS$ and $SPb$ at $y$=1.9-2.3,
the RQMD proton distributions tend to deviate from exponentials
in transverse mass, showing larger inverse slopes at low $m_{T}$ (within
the NA44 acceptance) than at high $m_{T}$. 
For RQMD, there is no significant 
difference in the inverse slopes if rope formation is turned off. This 
might seem surprising for antiprotons since if rope formation is included,
almost all antiprotons originate from ropes, and the production mechanisms
of antiprotons from ropes and from conventional sources are very different. 
However, rescattering would tend to 
mask any difference in the initial $m_T$ distribution of antiprotons
produced from either ropes or conventional sources.

The yields of protons and antiprotons both increase with system size, the
proton yield rising faster. 
The increase of both yields and inverse slopes 
with target size is much stronger in 
$SA$ than in $pA$ collisions
since target nucleons may be struck by more than one projectile nucleon.
The $\bar{p}/p$ ratio decreases with system size, and is lower towards
target rapidity than at mid-rapidity.
This implies that most of the protons at mid-rapidity are not produced in
the collision, but are remnants of the 
initial nuclei. The fraction and absolute numbers of such `original'
protons decreases from $y$=1.9-2.3 to $y$=2.3-2.9, indicating that the
stopping of protons at these energies is incomplete.
For $pA$ collisions, the antiproton yield is essentially constant from
$y$=1.9-2.3 to $y$=2.3-2.9, whereas for $SA$ collisions more antiprotons
are produced at $y$=2.3-2.9. 

For the $SA$ data, no change in the inverse slopes of protons and antiprotons
with centrality is seen within the 11\% most central $SPb$, and 
8\% most central $SS$, collisions. 
Both proton and antiproton yields increase with centrality.
 The increase is
larger for protons than for antiprotons, and both particles show a larger 
increase at $y$=2.3-2.9 than at $y$=1.9-2.3. 

RQMD predictions for proton and antiproton yields, with and without color 
ropes, are shown in Table~\ref{tb:rqmddndy} and Figure~\ref{fg:dnvrqmd}.
The proton yields are generally well described by the model
in both $pA$ and $SA$ collisions, although there is a tendency for RQMD to
overpredict the yield in the more backward rapidity interval for 
sulphur-nucleus collisions.
Antiproton yields 
from $pA$ collisions are also in reasonable
agreement with RQMD, however the data show a somewhat weaker target
dependence than the model.
The antiproton yield in $SS$ 
collisions is consistent with the RQMD prediction 
without ropes, 
whereas the $SPb$ yield is between the 
rope and no-rope predictions.
With no rope formation, it is possible that RQMD could reproduce the
observed $SA$ antiproton yields with a smaller annihilation 
cross-section in the model.
If, however, the rope hypothesis is valid, then
either RQMD overpredicts antiproton 
production in nucleus-nucleus collisions, or it underpredicts the
subsequent annihilation of the produced antiprotons in the surrounding medium.

Preliminary results from NA44 on the $\pi^+$ yields near mid-rapidity
($y$=3.0-4.0) show 
that the pion yield increases by a factor of 
36 $\pm$ 4
from $pBe$ to $SPb$ 
while the proton and antiproton yields increase by factors
of 75 $\pm$ 4 and 20 $\pm$ 2 respectively. 
The fact that antiproton production increases less rapidly with system size
than pion production naively suggests that the
antiproton yield from $SPb$ collisions may be lowered because of annihilation.

\begin{figure}
 \begin{center}\mbox{
     \epsfxsize=16cm
     \epsffile{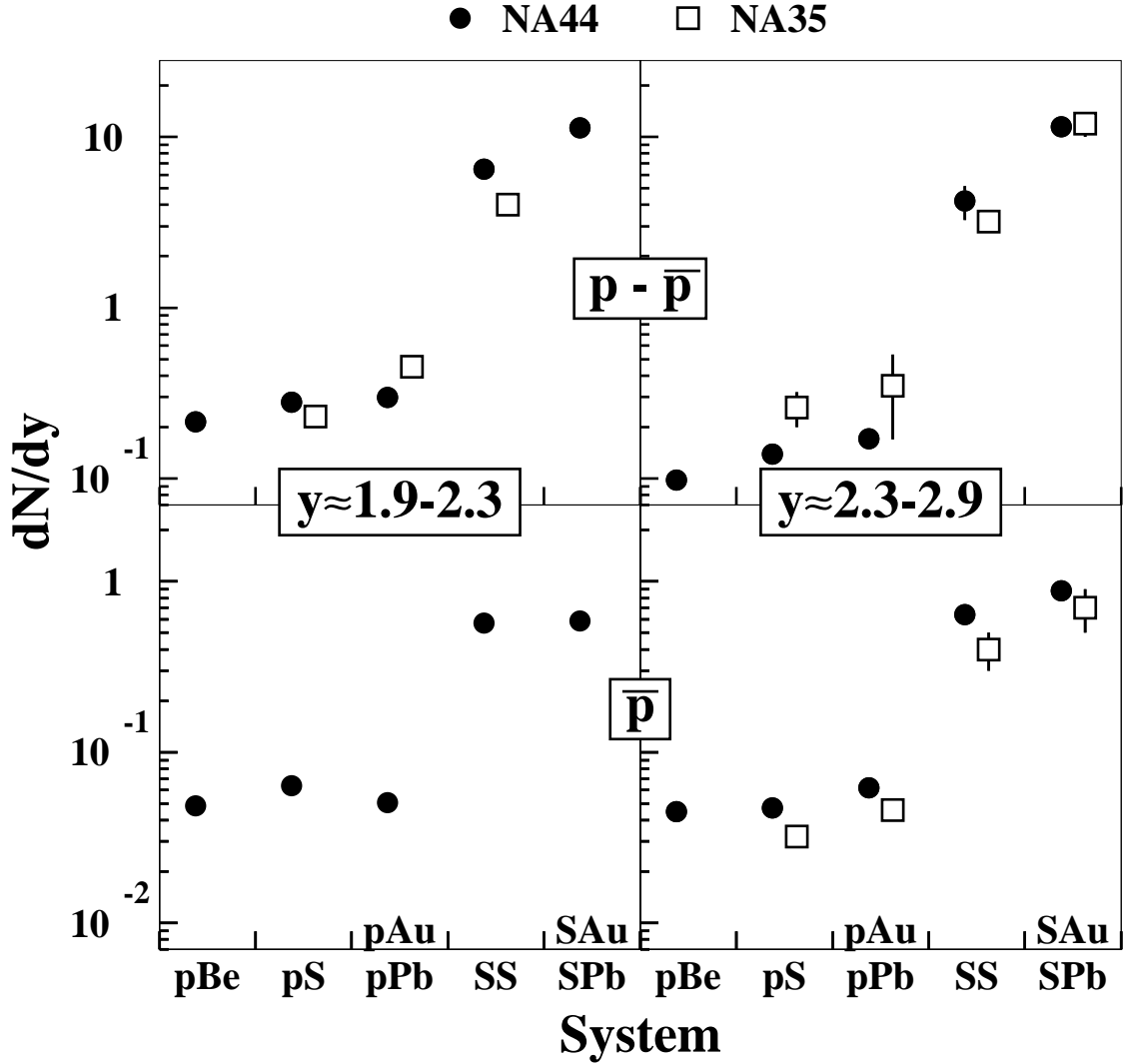}}\end{center}
  \caption{Comparison of NA44 and NA35, 
\protect{\cite{PROT35,PBAR35}}, 
 rapidity densities for
$p-\bar{p}$ and $\bar{p}$. 
 The $pA$ data were taken at 200 GeV/c for NA35 and
450 GeV/c for NA44. The NA44 $SS$ and $SPb$ data 
have been selected to have approximately the
same centrality selection as the NA35 $SS$ and $SAu$ data,
 and all NA44 data 
have been corrected for feed-down using the factors in
 Table~\protect{\ref{LAMCOR}}. The 
NA35 $p-\bar{p}$ data are from the rapidity intervals 
2.0-2.5 and 2.3-3.0 for both $SS$
and $SAu$; for $pA$ the NA35 rapidity ranges are 1.8-2.2 
and  the average of two bins covering rapidity 2.2-3.0.
The NA35 $\bar{p}$ data are in the rapidity
range 3-4.}
 \label{NA44V35}
\end{figure}

NA35 has published data on "net protons" (i.e. the difference between
protons and antiprotons) and antiprotons for $pS$, $pAu$, $SS$ (3\% most 
central) and $SAu$ (6\% most central) collisions 
\cite{PROT35,PBAR35}. In those cases
for which comparison data are available, the inverse slopes measured by
the two experiments are consistent.
Figure~\ref{NA44V35} 
shows a comparison of the
rapidity densities for net protons and antiprotons from NA44 and NA35.
There is good agreement between the two experiments.

\section*{Conclusions}
These data constitute the first
systematic measurement of proton and antiproton 
yields and spectra from $pBe$ to $SPb$. 
The stopping of protons is incomplete at 200GeV/A. 
For $SA$ collisions, target nucleons 
tend to be struck by more than one projectile nucleon: this causes the
target dependence of the inverse slopes and yields to be stronger
in $SA$ than in $pA$ collisions.
As the size of the 
system increases, the increasing density of particles at mid-rapidity causes
the protons to recscatter more often and so their mean $m_T$, or 
inverse slope, increases.
This mechanism is less efficient for antiprotons since they may annihilate
when they rescatter.
In spite of annihilation, the yield of antiprotons increases with system 
size and centrality:
this increase is strongest at mid-rapidity.
Comparisons with RQMD imply that
antiproton annihilation is overestimated in the model or that some new
mechanism is needed to account 
for antiproton production in sulphur-nucleus collisions.

\section{Acknowledgements}
NA44 is grateful to the staff of the CERN PS--SPS
accelerator complex for their excellent work. We thank the technical
staff at CERN and the collaborating institutes for their valuable
contributions. We are also grateful for the support given by the
 Austrian Fonds zur F\"orderung der Wissenschaftlichen Forschung (grant
P09586); the Science Research Council of Denmark; the Japanese Society
for the Promotion of Science; the Ministry of Education, Science and
Culture, Japan; the Science Research Council of Sweden; the US
W.M.~Keck Foundation; the US National Science Foundation; and the US
Department of Energy.

\begin{table}[htbp]
 \begin{center}\mbox{ }
   \begin{tabular}{|c|c|c|c|r|r|r|r|} \hline
{\bf System} &  {\bf Centrality}  & {\bf Angle}  &
 {\bf Target}  &
 \multicolumn{2}{c}{\bf $y$=1.9-2.3} &
 \multicolumn{2}{c|}{\bf $y$=2.3-2.9} \\
 &  & \bf{(mrad)} & \bf{Thickness} & 
 \multicolumn{1}{c|}{\bf $p$} & \multicolumn{1}{c|}{\bf $\bar{p}$} &
 \multicolumn{1}{c|}{\bf $p$} & \multicolumn{1}{c|}{\bf $\bar{p}$}  \\ \hline
$pBe$ & 84$\pm$2\% &  44 & 3.3\% & 22594 & 14925 & 35630 & 17880 \\ \cline{3-8}
     &                 & 131 & 3.3\% & 11200 &  3269 &  5568 &  1823 \\ \hline
$pS$  & 90$\pm$2\%  &  44 & 3.3\% & 44505 & 12081 &  1609 &  5392 \\ \cline{3-8}
     &                 & 131 & 3.3\% & 51341 &  1956 &       &       \\ \hline
$pPb$ & 97$\pm$3\%  &  44 & 4.7\% & 14094 & 13754 &  1913 &  4467 \\ \cline{3-8}
     &                 & 131 & 9.9\% & 22908 &   856 & 13907 &  2274 \\ \hline
$SS$  & 8.7$\pm$0.5\% &  44 & 6.6\% & 11135 &  3200 &  5960 & 16660 \\ \cline{3-8}
     &                 & 131 & 6.6\% & 18652 &  2703 & 18379 &  3762 \\ \hline
$SPb$ & 10.7$\pm$0.6\% & 44 & 5.9\% & 13833 &  2081 & 12062 &  7293 \\ \cline{3-8}
     &                 & 131 & 5.9\% & 38178 &  4528 & 43261 &  7299 \\ \hline
   \end{tabular}
  \end{center}
  \caption{Centrality, target thickness and 
number of events for each spectrometer setting. 
The target thickness is quoted in
nuclear collision lengths for the given system.}
  \label{tb:data}
\end{table}
\begin{table}[htbp]
 \begin{center}\mbox{ }
   \begin{tabular}{|c|c|c|c|} \hline
 {\bf System} &  \multicolumn{2}{c|}{\bf Error on Inverse Slope} 
 &  \multicolumn{1}{c|}{\bf Error on dN/dy} \\
 &    \multicolumn{1}{c|}{\bf $y$=1.9-2.3} & {\bf $y$=2.3-2.9} &  \\ \hline
  $pBe$ & 10 MeV/c & 10 MeV/c & 9\% \\ 
  $pS$  & 10 MeV/c & 10 MeV/c & 9\% \\ 
  $pPb$ & 10 MeV/c & 10 MeV/c & 10\% \\
  $SS$  & 20 MeV/c & 10 MeV/c & 9\% \\ 
  $SPb$ & 20 MeV/c & 10 MeV/c & 14\% \\ \hline
   \end{tabular}
  \end{center}
  \caption{Systematic errors on the inverse slopes and dN/dy.}
  \label{tb:errors}
\end{table}
\begin{table}[htbp]
 \begin{center}\mbox{ }
  \begin{tabular}{|c|c|c|c|c|c|} 
    \hline
{\bf System} & \multicolumn{1}{c|}{\bf Parameter} & \multicolumn{2}{c} {\bf $y$=1.9-2.3} & 
\multicolumn{2}{c|}{\bf $y$=2.3-2.9} \\ 
  &    & {\bf Proton} & {\bf Antiproton} & {\bf Proton} & {\bf Antiproton} \\ \hline
$pBe$  &     T&1.01$\pm$0.01&1.03$\pm$0.01&1.02$\pm$0.01&1.03$\pm$0.01\\ \cline{2-6}
       & dN/dy&0.90$\mp$0.03&0.81$\mp$0.04&0.90$\mp$0.03&0.85$\mp$0.04\\ \hline
$pS$   &     T&1.01$\pm$0.01&1.02$\pm$0.01&1.03$\pm$0.02&1.07$\pm$0.03\\ \cline{2-6}
       & dN/dy&0.91$\mp$0.03&0.79$\mp$0.05&0.91$\mp$0.04&0.83$\mp$0.06\\ \hline
$pPb$  &     T&1.09$\pm$0.02&1.29$\pm$0.24&1.04$\pm$0.01&1.11$\pm$0.03\\ \cline{2-6}
       & dN/dy&0.82$\mp$0.04&0.72$\mp$0.06&0.93$\mp$0.02&0.81$\mp$0.04\\ \hline
$SS$   &     T&1.05$\pm$0.01&1.09$\pm$0.03&1.02$\pm$0.01&1.08$\pm$0.02\\ \cline{2-6}
       & dN/dy&0.78$\mp$0.05&0.74$\mp$0.06&0.84$\mp$0.04&0.71$\mp$0.06\\ \hline
$SPb$  &     T&1.01$\pm$0.01&1.09$\pm$0.02&1.10$\pm$0.02&1.12$\pm$0.03\\ \cline{2-6}
       & dN/dy&0.77$\mp$0.06&0.67$\mp$0.07&0.82$\mp$0.05&0.73$\mp$0.06\\ \hline
  \end{tabular}
  \end{center}
  \caption{`Feed-down factors' for $\Lambda$ and $\Sigma$ decays.
$T$ is the inverse slope of the $m_T$ spectrum. 
These are multiplicative factors which could be applied to the numbers in
Tables~\protect{\ref{tb:slope}} and \protect{\ref{tb:dndy}} to estimate the 
inverse slopes 
and yields of `direct' protons (antiprotons). The errors reflect the 
result of increasing and decreasing the $(\Lambda +\Sigma)/p$ ratio in RQMD 
by a factor of 1.5. The errors on T and dN/dy are anticorrelated.}
  \label{LAMCOR}
\end{table}
\begin{table}[htbp]
 \begin{center}\mbox{ }
   \begin{tabular}{|c|c|c|c|c|c|c|} \hline
{\bf y} & {\bf Fit Range} & {\bf System} & {\bf p} & 
{\bf RQMD p } & {\bf $\overline p$ }  & {\bf RQMD $\overline p$} \\ 
 &  (GeV/c)        &  & (MeV/c) & (MeV/c) & (MeV/c) & (MeV/c) \\ \hline 
 &                 & $pBe$ &$123~\pm$~~  4 &$135~\pm$~~  5 &$116~\pm$   12 &$130~\pm$   13 \\
 &                 & $pS $ &$136~\pm$~~  3 &$146~\pm$~~  8 &$149~\pm$   25 &$130~\pm$   19 \\
 1.9~-~2.3 & $m_T-m_p \le 0.27$ & $pPb$ &$131~\pm$~~  3 &$146~\pm$~~  9 &$126~\pm$  16 &$126~\pm$   25 \\
 &                 & $SS $&$149~\pm$ ~~4 &$220~\pm$ 28 &$217~\pm$ 28 &$156~\pm$ 42 \\
 &                 & $SPb$&$195~\pm$ ~~5 &$270~\pm$ 14 &$207~\pm$ 17 &$187~\pm$ 26 \\ \hline
 &                 & $pBe$&$153~\pm$~~  4 &$150~\pm$~~  3 &$126~\pm$~~  6 &$142~\pm$  ~~5 \\
 &                 & $pS $&$170~\pm$   30 &$148~\pm$~~  4 &$131~\pm$~  10 &$143~\pm$  ~~9 \\
 2.3~-~2.9& $m_T-m_p \le 0.68$ & $pPb$ &$195~\pm$~~  5 &$157~\pm$~~  5 &$141~\pm$~~  8 &$140~\pm$   11 \\
 &                 & $SS $&$210~\pm$ ~~4 &$238~\pm$ ~~9 &$190~\pm$ ~10 &$171~\pm$ 15 \\
 &                 & $SPb$&$256~\pm$ ~~4 &$246~\pm$ ~~4 &$205~\pm$ ~~7 &$202~\pm$ ~~8 \\ \hline
   \end{tabular}
  \end{center}
  \caption{Inverse slopes (T) extracted from fits of the data to 
Equation~\protect{\ref{eq:mt}}. The errors are statistical. 
Systematic errors are shown in Table~\protect{\ref{tb:errors}}.
Also shown are the inverse slopes extracted from RQMD, version 1.08, after
correction for weak decay feed-down.
 For RQMD, there is no significant 
difference in the inverse slopes if rope formation is turned off.}
  \label{tb:slope}
\end{table}
\begin{table}[htbp]
 \begin{center}\mbox{ }
   \begin{tabular}{|c|c|c|c|} \hline

{\bf y}& \multicolumn{1}{c|} {\bf System} & {\bf Protons}          & {\bf Antiprotons} \\ \hline
        & $pBe$& $ 0.293 \pm 0.008\pm 0.026$ & $ 0.060 \pm 0.003\pm 0.005$ \\
        & $pS $& $ 0.377 \pm 0.009\pm 0.034$ & $ 0.080 \pm 0.008\pm 0.007$ \\
1.9~-~2.3 & $pPb$& $ 0.426 \pm 0.010\pm 0.043$ & $ 0.070 \pm 0.006\pm 0.007$ \\
        & $SS $& $~~~ 4.97  \pm 0.15~\: \pm 0.46~~~ $ & $ 0.413 \pm 0.036\pm 0.038$ \\
        & $SPb$& $~~~13.6   \pm 0.4~~~  \pm 1.9\,~~~~  $ & $ 0.791 \pm 0.043\pm 0.108$ \\ \hline
        & $pBe$& $ 0.158 \pm 0.008\pm 0.014$ & $ 0.053 \pm 0.006\pm 0.005$ \\
        & $pS $& $ 0.204 \pm 0.013\pm 0.018$ & $ 0.057 \pm 0.006\pm 0.005$ \\
2.3~-~2.9 & $pPb$& $ 0.251 \pm 0.010\pm 0.025$ & $ 0.076 \pm 0.007\pm 0.008$ \\
        & $SS $& $\:~~4.51  \pm 0.20~\, \pm 0.42~~~ $ & $ 0.505 \pm 0.038\pm 0.047$ \\
        & $SPb$& $~~~12.0   \pm 0.3~~\:  \pm 1.6\,~~~~  $ & $ 1.10  \pm 0.05~\, \pm 0.15 $ \\ \hline
   \end{tabular}
 \end{center}
  \caption{dN/dy for protons and antiprotons with statistical and 
 systematic errors.}
  \label{tb:dndy}
\end{table}

\begin{table}[htbp]
 \begin{center}\mbox{ }
   \begin{tabular}{|c|c|c|c|c|c|} \hline
{\bf y}&  {\bf System} & \multicolumn{2}{c}{\bf Ropes} &
 \multicolumn{2}{c|}{\bf No Ropes} \\
 & &   {\bf Protons} & {\bf Antiprotons}  & {\bf Protons} & {\bf Antiprotons} \\ \hline
          & $pBe$ &$  0.211 \pm 0.005$ &$  0.033 \pm 0.002$ &&\\
          & $pS $ &$  0.279 \pm 0.010$ &$  0.039 \pm 0.004$ &&\\
1.9~-~2.3 & $pPb$ &$  0.433 \pm 0.019$ &$  0.057 \pm 0.007$ &&\\
          & $SS $ &$ ~5.64  \pm 0.56~$ &$  0.712 \pm 0.122$ &$  6.19\pm 0.94$ &$  0.221 \pm 0.065$\\
          & $SPb$ &$ 16.66   \pm 0.73~  $&$1.050 \pm 0.105$ &$ 16.2 \pm 1.2~\,$ &$  0.355 \pm 0.163$\\ \hline
          & $pBe$ &$  0.179 \pm 0.003$ &$  0.048 \pm 0.002$ &&\\
          & $pS $ &$  0.221 \pm 0.005$ &$  0.058 \pm 0.003$ &&\\
2.3~-~2.9 & $pPb$ &$  0.295 \pm 0.009$ &$  0.095 \pm 0.005$ &&\\
          & $SS $ &$  5.252 \pm 0.155$ &$  1.051 \pm 0.073$ &$  4.98\pm 0.24$ &$  0.366 \pm 0.064$\\
          & $SPb$ &$ 12.71  \pm 0.13~$ &$  1.589 \pm 0.049$ &$ 11.22  \pm 0.18~\, $ &$  0.462 \pm 0.037$\\ \hline
   \end{tabular}
 \end{center}
  \caption{Predictions for dN/dy for protons and antiprotons from
RQMD, version 1.08. The contribution from weak decay feed-down is included in 
these calculations.}
  \label{tb:rqmddndy}
\end{table}
\end{document}